\documentclass[aps,pra,floatfix,showpacs,twocolumn]{revtex4}
\usepackage{amsmath}
\usepackage{graphicx}
\def\prn#1{{\left(#1\right)}}

\newcommand{\abs}[1]{\left|#1\right|}

\def\bra#1{{\left\langle#1\right\vert}}
\def\ket#1{{\left\vert#1\right\rangle}}

\def\prn#1{{\left(#1\right)}}
\def\brk#1{{\left[#1\right]}}

\def\abs#1{{\left|#1\right|}}

\newcommand{\oberlin}{Department of Physics and Astronomy, Oberlin College, Oberlin, OH 44074, USA}

\begin{document}
\title{Velocity-selective direct frequency-comb spectroscopy of atomic vapors}
\author{J.\ E.\ Stalnaker}\email{jason.stalnaker@oberlin.edu}
\author{S.\ L.\ Chen}
\author{M.\ E.\ Rowan}
\affiliation{\oberlin}

\author{K.\ Nguyen}
\affiliation{Department of Physics, San Jos\'e State University, San Jos\'e, California 95192, USA}

\author{T.\ Pradhananga}
\affiliation{Department of Physics, California State University -- East Bay, Hayward, California 94542-3084, USA}

\author{C.\ A.\ Palm}
\affiliation{Department of Physics, California State University -- East Bay, Hayward, California 94542-3084, USA}

\author{D.\ F.\ Jackson Kimball}
\email{derek.jacksonkimball@csueastbay.edu}
\affiliation{Department of Physics, California State University -- East Bay, Hayward, California 94542-3084, USA}

\date{\today}
\begin{abstract}
We present an experimental and theoretical investigation of two-photon direct frequency-comb spectroscopy performed through velocity-selective excitation.  In particular, we explore the effect of repetition rate on the $\textrm{5S}_{1/2}\rightarrow \textrm{5D}_{3/2, 5/2}$ two-photon transitions excited in a rubidium atomic vapor cell.  The transitions occur via step-wise excitation through the $\textrm{5P}_{1/2, 3/2}$ states by use of the direct output of an optical frequency comb.  Experiments were performed with two different frequency combs, one with a repetition rate of $\approx 925$~MHz and one with a repetition rate of $\approx 250$~MHz.  The experimental spectra are compared to each other and to a theoretical model.
\end{abstract}


\maketitle
\section{Introduction}

The use of ultra-short pulses to perform precision spectroscopy has gained increased interest due to the development of phase-stabilized optical frequency combs (see Ref.\ \cite{stowe08} for a review of direct frequency-comb spectroscopy).  One of the chief advantages of using the output of an optical frequency comb to directly excite atomic transitions is the spectral versatility afforded by the comb.  Stabilized optical frequency combs of the types used here contain $\sim 10^5 - 10^6$ optical frequencies spanning the entire visible spectrum.  Thus, the optical frequency comb acts effectively as a large number of cw lasers with frequencies given by
\begin{align}
\omega_n = 2\pi\prn{n\, f_r+f_0}, \label{eq combEq}
\end{align}
where $n$ is an integer denoting the mode of the comb, $f_r$ is the repetition rate of the laser, and $f_0$ is the carrier-envelope-offset frequency (for a review of frequency combs see, e.g., Ref.\ \cite{cundiff03}).

Two-photon direct frequency-comb spectroscopy with a fully stabilized comb was first applied to the study of cold rubidium atoms \cite{marian04,marian05,stowe06}. In these experiments independent control of $f_0$ and $f_r$ was used to explore the effects of resonant enhancement of the two-photon transition rate due to resonance with an intermediate state.  Two-photon direct frequency-comb spectroscopy has also been applied to the study of room-temperature atomic cesium confined to a vapor cell \cite{stalnaker10}.  The transitions studied in that work were excited via two-photon excitation through a resonant intermediate state using velocity-selective excitation.

Here, we further explore velocity-selective, two-photon excitation both theoretically and experimentally.  In particular, we consider the effect of the repetition rate of the comb on the spectra.  We identify two different regimes: one in which the repetition rate of the comb is larger than the Doppler width of the resonance corresponding to the first stage of the transition and one in which the repetition rate is smaller than the Doppler width of the resonance of the first stage of the transition.   We present data for the $\textrm{5S}_{1/2} \rightarrow \textrm{5P}_{1/2} \rightarrow \textrm{5D}_{3/2}$ and $\textrm{5S}_{1/2} \rightarrow \textrm{5P}_{3/2} \rightarrow \textrm{5D}_{3/2,5/2}$ transitions in atomic rubidium taken with two different optical frequency combs with different repetition rates.  We compare our results to a model calculation and see how the velocity selection process differs for the two cases.  In addition, we explore how the energy of the intermediate state affects the two-photon spectra and explain the qualitative differences in the spectra that arise from excitation through different intermediate states.

\section{Theoretical Considerations}

\subsection{Velocity Selective Resonance}
The two-photon transition probability between two states $\ket{g}$ and $\ket{f}$ is given by \cite{loudin83}

\begin{widetext}
\begin{align}
W\prn{g ,\, f} = & \prn{\frac{ I_1\, I_2}{4\, \epsilon_0^2\, c^2\, \hbar^4}} \frac{\gamma_f}{\brk{\omega_{g:f}-\prn{\omega_1+\mathbf{k}_1\cdot\mathbf{v}}-\prn{\omega_2+\mathbf{k}_2\cdot \mathbf{v}}}^2+\prn{\frac{\gamma_f}{2}}^2} \times \abs{\sum_k\frac{\bra{f}\hat{e}_2\cdot\mathbf{d}\ket{k}\bra{k}\hat{e}_1\cdot\mathbf{d} \ket{g}}{\omega_{g:k}-\prn{\omega_1+\mathbf{k}_1\cdot\mathbf{v}}-i\,\frac{\gamma_{k}}{2}}}^2
, \label{eq twoPhotonProb}
\end{align}
\end{widetext}
where $\mathbf{v}$ is the velocity of the atom, $\mathbf{d}$ is the electric dipole operator, $\hat{e}_{1(2)}$ is the unit vector along the direction of the
polarization for the first(second) laser beam of intensity $I_{1(2)}$, $\gamma_{k}$ and $\gamma_{f}$ are the homogeneous line widths of the states $\ket{k}$ and $\ket{f}$, respectively, $\omega_{g:k}$ and $\omega_{g:f}$ are the resonant angular frequencies of the transitions $\ket{g} \rightarrow \ket{k}$ and $\ket{g} \rightarrow \ket{f}$, respectively.  The sum over $k$ includes all possible intermediate states.  We have assumed that the laser frequency $\omega_1$ is near the single-photon resonance of the $\ket{g} \rightarrow \ket{k}$ transitions and have neglected the possibility of $\omega_2$ exciting the first stage of the transition.  If we further assume that there is a single intermediate state that is close in energy to $\omega_1$ we can reduce the sum to a single term consisting of the nearly resonant intermediate state.  The maximum transition probability will occur when both the two-photon resonance condition and the single-photon resonance condition are met:
\begin{align}
  \omega_{g:f} = & \omega_1+\mathbf{k}_1\cdot\mathbf{v}+\omega_2+\mathbf{k}_2\cdot \mathbf{v} \label{eq twoPhotonRes}\\
  \omega_{g:k} = & \omega_1+\mathbf{k}_1\cdot\mathbf{v} . \label{eq onePhotonRes}
\end{align}

When an optical frequency comb is used to excite an atomic transition, the transition rate must be summed over all of the optical modes in the comb [the frequencies of which are given by Eq.\ \eqref{eq combEq}].  Resonance will occur if there are any comb modes  $n_1$ and $n_2$ resulting in frequencies $\omega_1$ and $\omega_2$ that simultaneously satisfy Eqs.\ \eqref{eq twoPhotonRes} and \eqref{eq onePhotonRes}.  If we consider the situation where the comb light is counter-propagated through the atomic sample (so that $\hat{k}_1 = -\hat{k}_2$) we can write the resonance conditions as
\begin{align}
  \frac{\omega_{g:f}}{2\pi} = & \prn{n_1+n_2}\, f_r + \prn{n_1-n_2}\frac{v_x}{c}\, f_r +2 f_0 \label{eq combTwoPhotonRes}\\
  \frac{\omega_{g:k}}{2\pi} = & \prn{n_1\, f_r +f_0}\prn{1+\frac{v_x}{c}}   , \label{eq combOnePhotonRes}
\end{align}
where we have defined the $x$ axis to be the direction of propagation of the light and taken $v_x$ to be the $x$ component of the atomic velocity.  For a given $n_1$, $n_2$, and $f_0$ there exists a repetition rate, $f_r$, which will satisfy the above equations for some value of $v_x$.  If the velocity class corresponding to $v_x$ that satisfies the resonance conditions is populated in the atomic distribution, there will be resonant excitation to the state $\ket{f}$ in the atomic sample.

In this work we have investigated this velocity-selective excitation with two different frequency combs having different repetition rates.  The first situation we consider is the case where the repetition rate of the comb is larger than the Doppler width of the resonance corresponding to the first stage of the excitation.  In this case there will exist, at most, one comb mode $n_1$ which is resonant with a velocity class present in the atomic sample.  As a result, there will be at most a single pair of comb modes $n_1$ and $n_2$ that sum up to a total mode number $n_T=n_1+n_2$ that satisfy the resonance conditions given by Eqs.\ \eqref{eq combTwoPhotonRes} and \eqref{eq combOnePhotonRes} for a velocity class present in the atomic sample.  As the repetition rate is scanned the velocity class excited to the first stage of the transition changes as the frequency $\omega_1$ sweeps over the Doppler broadened transition.  Once the velocity class that is excited to the intermediate state $\ket{k}$ also satisfies the resonance condition for $\omega_2 = 2\pi\prn{n_2\, f_r+f_0}$, the atoms will be excited to the final state $\ket{f}$. This results in a sub-Doppler resonant peak.

If the repetition rate is increased over a large enough range a different pair of comb modes, adding up to $n_T^\prime = n_1+n_2-1$ will satisfy the resonance requirement and the resonant peak will recur.  However, because the velocity class that satisfies the resonant condition will, in general, be different, the relative amplitude of the $n_T^\prime$ peak may vary greatly as compared to the $n_T$ peak; reflecting the relative population in the resonant velocity classes.  The amount the repetition rate must scan before the spectrum repeats is
\begin{align}
  \Delta f_r \approx 2\pi\, \frac{f_r^2}{\omega_{g:f}} . \label{eq repInterval}
\end{align}
If the laser beams are exactly counter-propagating throughout the excitation region then the width of the resonance in terms of the repetition rate, $\delta f_r$, is approximately the natural line width of the excited state divided by the total mode number:
\begin{align}
  \delta f_r \approx \frac{\gamma_f}{2\, \pi\prn{n_1+n_2}} . \label{eq fRWidth}
\end{align}

\begin{figure}[h]
\centering
\includegraphics[width=3.375in]{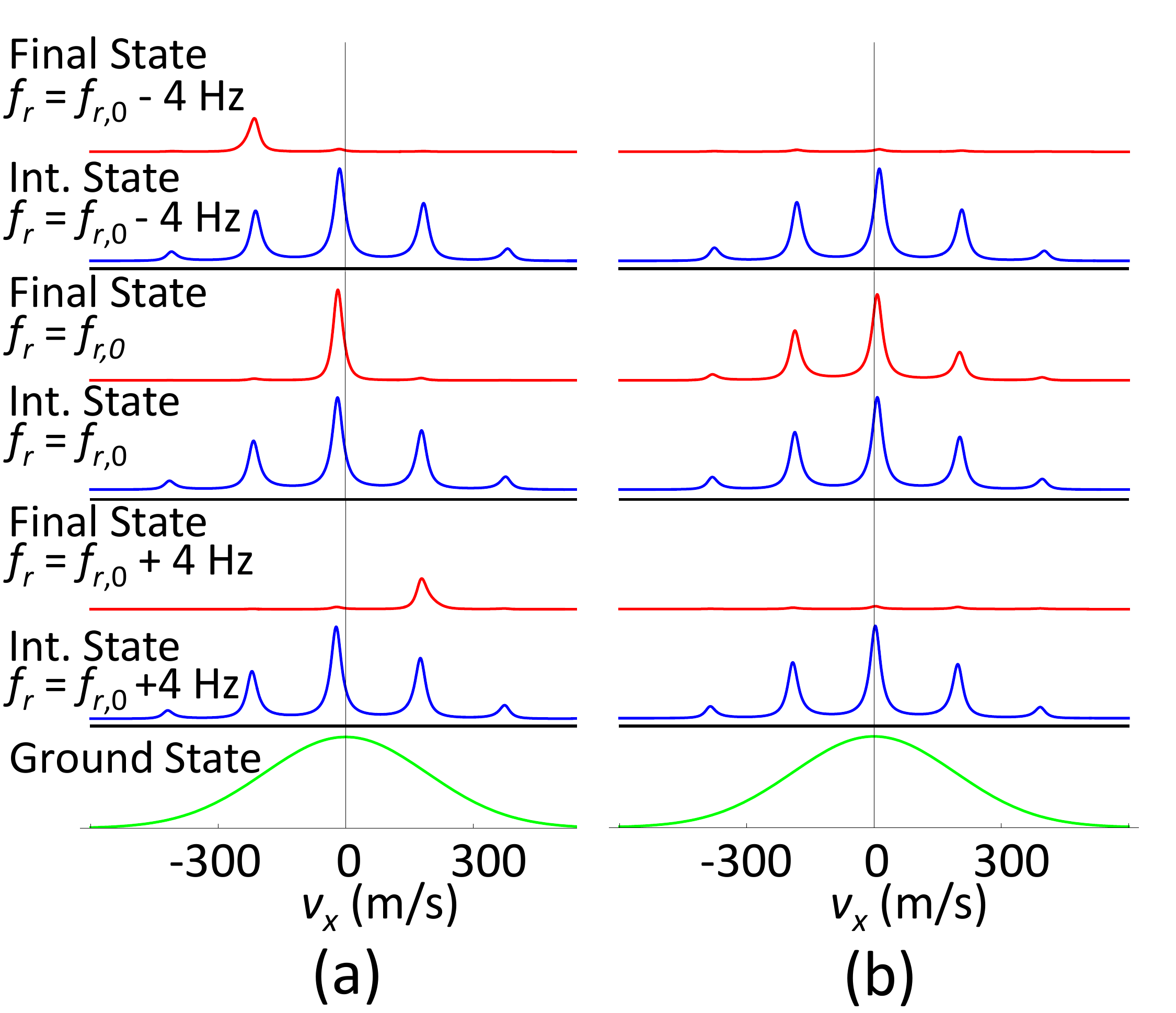}
\caption{Population of the intermediate state and excited state as a function of velocity for three different values of the repetition rate.  The bottom trace shows the ground state velocity distribution for a rubidium vapor cell at a temperature of $T= 373$ K.  As the repetition rate is varied the velocity classes excited to the intermediate state move through the distribution.  When the repetition rate satisfies the resonant condition for a given ($n_1$, $n_2$) and for a specific velocity class atoms are excited to the final state.  Figure (a) shows the case where the resonant repetition rate is different for the different mode pairs and velocity classes, leading to multiple peaks in the spectrum.   The calculation is based on a three-level model where the frequencies of the intermediate and final states correspond to the frequencies of the $\textrm{5D}_{3/2}$ state excited through the $\textrm{5P}_{1/2}$ state.  Figure (b) shows the case where the intermediate state is close to the halfway point of the energy between the ground and final states.  In this case, the resonant repetition rate will be the same for the different velocity classes and different mode numbers that have the same $n_T=n_1+n_2$.  The frequencies used in the calculation correspond to the frequencies of the $\textrm{5D}_{5/2}$ state excited through the $\textrm{5P}_{3/2}$ state.  The parameters used for the comb light correspond to those used in the experiment with the smaller repetition rate frequency comb ($f_r \approx 250$ MHz). }\label{fig vSelectionSmallfR}
\end{figure}

The second case we consider is that in which the repetition rate of the frequency comb is less than the Doppler width of the resonance corresponding to the first stage of the transition. In this case there are multiple comb modes resonant with the first stage of the transition for different velocity classes.  Consequently, there are multiple pairs of comb modes ($n_1,\, n_2$), having the same total mode number $n_T = n_1+n_2$, that will lead to resonant excitation for different velocity classes.  In general, the different velocity classes will be resonant at different repetition rates.  The result is a spectrum that has a comb-like structure as different pairs of comb modes contribute.  Figure \ref{fig vSelectionSmallfR}~(a) shows the excitation of the different velocity classes to the intermediate and final states for three different values of the repetition rate.  The resonant repetition rate, $f_{r,0}$, is the repetition rate at which the largest number of atoms are are excited to the final state.  This corresponds to excitation of a velocity class that is near zero.  There are additional resonances when the repetition rate is below (above) the resonance value, leading to excitation of velocity classes that have negative (positive) velocity along the propagation direction of the laser beam that excites the first stage of the transition.

\begin{figure}[h]
\centering
\includegraphics[width=3.375in]{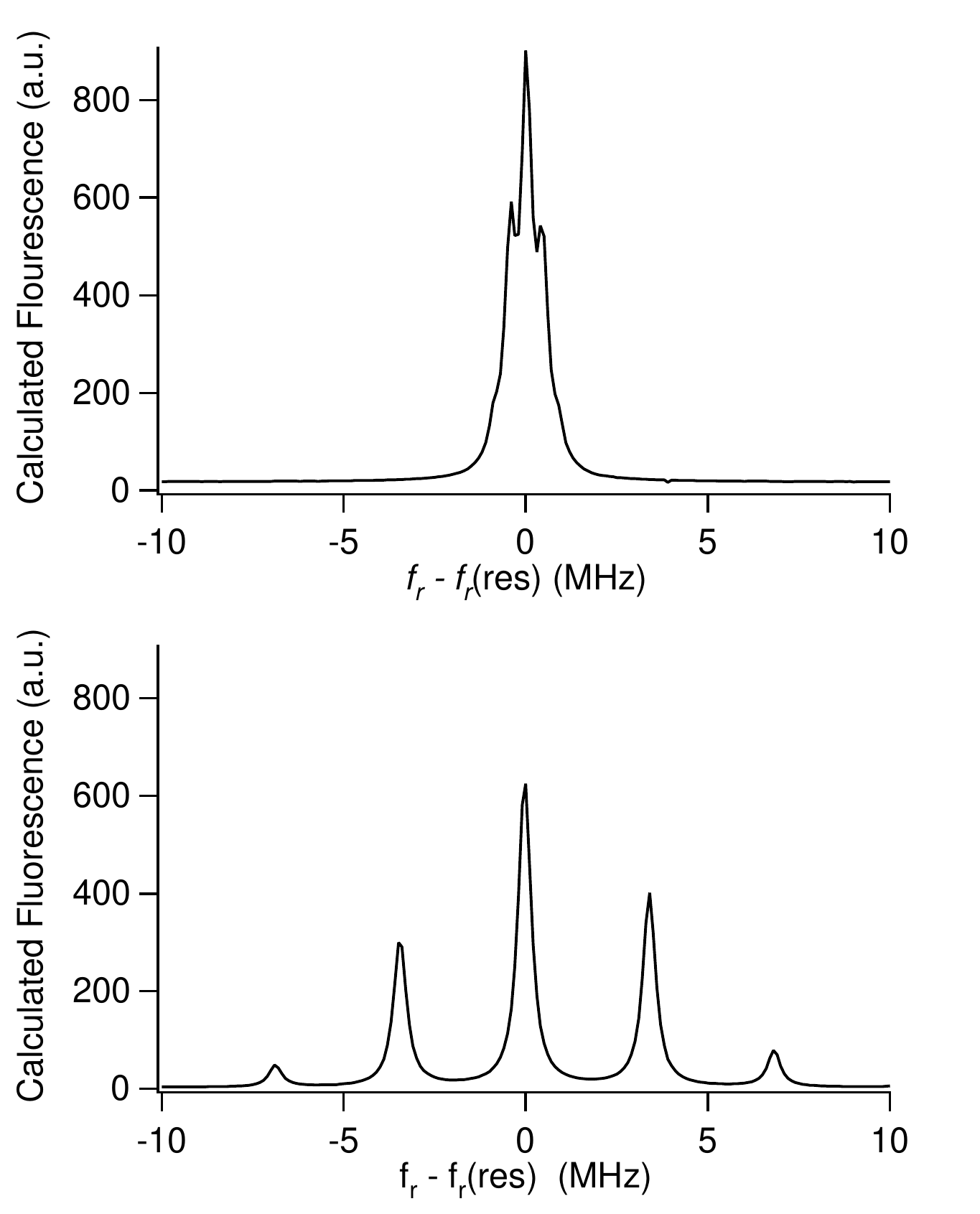}
\caption{Calculated fluorescence spectra for a three-level system.  The top plot shows the calculated spectrum for the situation where the intermediate state is close to half of the two-photon transition energy (1.5~THz from the mid point).  The energies used correspond to those of the $\textrm{5S}_{1/2}\rightarrow \textrm{5P}_{3/2} \rightarrow \textrm{5D}_{5/2}$ transition.  The different velocity classes are all resonant near the same repetition rate, resulting in a single peak, consisting of unresolved contributions from different velocity classes.  The bottom plot shows the calculated spectrum for the situation where the intermediate state is far from the mid point of the two-photon transition energy (7.5~THz from the mid point).  The energies used correspond to those of the $\textrm{5S}_{1/2}\rightarrow \textrm{5P}_{1/2} \rightarrow \textrm{5D}_{3/2}$ transition.  The parameters used for the comb light correspond to those used in the experiment with the small repetition rate frequency comb ($f_r \approx 250$~MHz). }\label{fig smallfRSpectra}
\end{figure}

The case where the repetition rate of the comb is less than the Doppler width of the first stage of the transition exhibits a significant change if the energy of the intermediate state is nearly halfway between the final and ground states' energy difference.  This is the case for the $\textrm{5S}_{1/2}\rightarrow \textrm{5P}_{3/2} \rightarrow \textrm{5D}_{5/2}$ transition in rubidium, where the intermediate state is only 1.5 THz from the mid-point of the ground-to-final-state energy separation.  In this case, the different velocity classes satisfy the resonance conditions at nearly the same repetition rates.  This can be understood by noting that the Doppler shift is wavelength dependent.  As a result, the apparent spacing of comb modes in ``velocity space" is similar if the wavelengths for the first and second stages of the transition are similar and are different for the two stages if the wavelengths are different.  In the case of where the wavelengths are similar the spectrum is significantly simpler than the more general case described above and there is no comb-like structure in the spectrum.   This situation is illustrated in Fig.\ \ref{fig vSelectionSmallfR} (b), where the energy levels were based on the $\textrm{5S}_{1/2}\rightarrow \textrm{5P}_{3/2} \rightarrow \textrm{5D}_{5/2}$ transition in rubidium.  The resonant peak contains contributions from multiple velocity classes that are simultaneously excited.  Figure \ref{fig smallfRSpectra} shows the calculated fluorescence spectra as the repetition rate is scanned for the three-level model both for the case where the energy of the intermediate state is nearly half that of the two-photon energy (1.5 THz away) and for the case where the intermediate state is not close to the half of the two-photon energy (7.5 THz away).  These values correspond to the $\textrm{5S}_{1/2}\rightarrow \textrm{5P}_{3/2} \rightarrow \textrm{5D}_{5/2}$ and $\textrm{5S}_{1/2}\rightarrow \textrm{5P}_{1/2} \rightarrow \textrm{5D}_{3/2}$ transitions, respectively.

This dependence of the character of the spectrum on the intermediate state energy can be understood by considering two pairs of modes $(n_1,\,  n_2)$ and $(n_1^\prime, \, n_2^\prime) = (n_1+1,\, n_2-1)$ that sum up to the same total mode number $n_T$.  We suppose that $f_r$ is the resonant repetition rate for a velocity class $v_x$ with mode numbers $(n_1, \, n_2)$ and $f_r^\prime$ is the resonant repetition rate for a different velocity class $v_x^\prime$ excited by frequencies with mode numbers $(n_1^\prime, \, n_2^\prime)$.  The condition that the two resonant peaks overlap is that $\Delta f_r = f_r^\prime - f_r < \frac{\gamma_f}{2\pi\, n_T}$.  Solving the resonance conditions [Eqs.\ \eqref{eq combTwoPhotonRes} and \eqref{eq combOnePhotonRes}] gives
\begin{align}
  \Delta f_r \approx \prn{\frac{n_1-n_2}{n_1\, n_2}} \frac{f_r}{2} , \label{eq resRequirement}
\end{align}
where we have kept only terms first order in $\Delta f_r$ and have assumed $f_r \gg f_0$ for simplicity.  In the case of the excitation of the $\textrm{5S}_{1/2}\rightarrow \textrm{5P}_{3/2} \rightarrow \textrm{5D}_{5/2}$ transition with a comb that has $f_r \approx 250$~MHz we have $n_1\approx 1.536 \times 10^6$ and $n_2\approx 1.544 \times 10^6$, giving $\Delta f_r \approx 0.4$ Hz.  This is smaller than the resonant line width of $\delta f_r \approx 0.9$, where we have used $\frac{\gamma_f}{2\pi} \approx 3$~MHz for the homogeneous line width of the transition.

\subsection{Two-Photon Transitions in Rubidium}
\begin{figure}[h]
\centering
\includegraphics[width=3.375in]{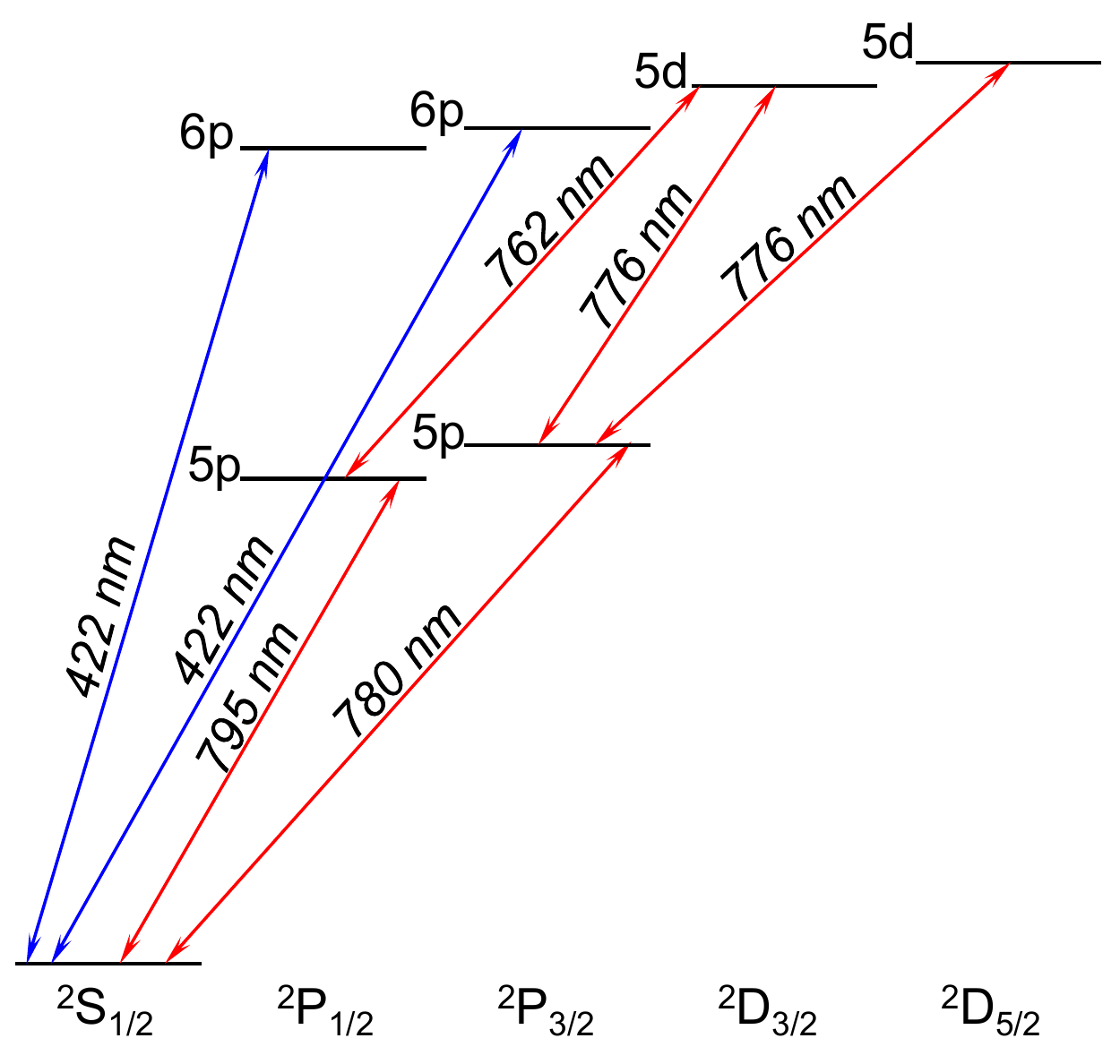}
\caption{Energy level diagram for Rb showing the relevant energy levels. }\label{fig RbEnergy}
\end{figure}

Figure \ref{fig RbEnergy} shows an energy level diagram for the two stable isotopes of rubidium ($^{85}\textrm{Rb}$ abundance $72.2$\% and $^{87}\textrm{Rb}$ abundance $27.8$\%). The transitions considered here are transitions to the $\textrm{5D}_{5/2}$ and $\textrm{5D}_{3/2}$ states through the $\textrm{5P}_{1/2}$ and $\textrm{5P}_{3/2}$ intermediate states.  The nonzero nuclear spin of both isotopes ($I = 5/2$ for $^{85}\textrm{Rb}$ and $I = 3/2$ for $^{87}\textrm{Rb}$) result in hyperfine structure for all of the states involved.  Thus, there are a number of possible transitions leading to a complex and rich fluorescence spectrum.  Including the hyperfine structure, the general two-photon transition rate becomes
\begin{widetext}
\begin{small}
\begin{align}
W\prn{5\textrm{S}_{1/2} F ,\, 5\textrm{D}_{J^{\prime\prime}} F^{\prime\prime}} = & \prn{\frac{I_1\, I_2}{4\,\epsilon_0^2\, c^2\, \hbar^4}}\prn{\frac{1}{2F+1}} \frac{\gamma_{5\textrm{D}_{J^{\prime\prime}}}}
{\brk{\omega_{5\textrm{S}_{1/2} F:5\textrm{D}_{J^{\prime\prime}}F^{\prime\prime}}-\prn{\omega_1+\mathbf{k}_1\cdot\mathbf{v}}-\prn{\omega_2+\mathbf{k}_2\cdot \mathbf{v}}}^2+\prn{\frac{\gamma_{5\textrm{D}_{J^{\prime\prime}}}}{2}}^2} \nonumber \\
&\;\;\;\;\;\;\;\;\;\;\;\; \times \sum_{M_F,\, M_F^{\prime\prime}}
\abs{\sum_{J^\prime ,F^\prime ,M_F^\prime}\frac{\bra{5\textrm{D}_{J^{\prime\prime}}F^{\prime\prime}
M_F^{\prime\prime}}\hat{e}_2\cdot\mathbf{d}\ket{5\textrm{P}_{J^\prime}F^\prime
M_F^\prime}\bra{5\textrm{P}_{J^\prime}F^\prime
M_F^\prime}\hat{e}_1\cdot\mathbf{d}\ket{5\textrm{S}_{1/2}F
M_F}}{\omega_{5\textrm{S}_{1/2} F: 5 \textrm{P}_{J^{\prime}}
F^{\prime}}-\prn{\omega_1+\mathbf{k}_1\cdot\mathbf{v}}-i\,\frac{\gamma_{5\textrm{P}_{J^\prime}}}{2}}}^2
, \label{eq twoPhotonProb}
\end{align}
\end{small}
\end{widetext}
where $M_F$, $M_F^\prime$,
and $M_F^{\prime\prime}$ are the projections of the total angular
momenta $F$, $F^\prime$, and $F^{\prime\prime}$ along the axis of
quantization, $\gamma_{n \, L_J}$
is the homogeneous line width of the state $\ket{n \, L_J}$, and
$\omega_{n\, L_{J} F: n^{\prime}\, L^{\prime}_{J^{\prime}}F^\prime}$
is the resonant angular frequency of the transition $\ket{n\, L_{J}
\, F} \rightarrow \ket{ n^{\prime}\, L^{\prime}_{J^{\prime}}\,
F^\prime}$.  The sum over $J^\prime$ runs from $1/2$ to $3/2$, and the $F^\prime$ runs over the hyperfine levels of the intermediate state.  For given polarizations of the two beams, the Wigner-Eckart theorem can be used to relate the matrix elements to a reduced matrix element that is independent of the magnetic sublevels.  The reduced matrix elements in the $F$ basis can be related to the reduced matrix elements in the $J$ basis using standard angular momentum relations (see, e.g., \cite{sobelman96}).  Using these relations, the spectra were calculated by inputting known values for the transition frequencies \cite{barwood91,ye96,banerjee04,nez93} and integrating the transition probability over the velocity distribution.  The number of comb modes used for the calculation varied depending on the repetition rate being considered.  Typically, 15-20 modes were used in the calculations that were done for the high repetition rate laser ($f_r \approx 924$ MHz) and 20-25 modes were used for the calculations for the lower repetition rate laser ($f_r \approx 250$~MHz).

The calculation also accounted for depletion of the power for the first stage of the transition as the beam propagated through the vapor cell.  The two-photon transition rate is proportional to the product of the intensities of the resonant laser fields.  As a result it is advantageous to focus the light in order to increase the transition rate.  This results in a localized region that provides the largest contribution to the fluorescence signal.  However, if the atomic density is large enough, the power resonant with the first stage of the transition can be depleted through absorption before the focused region of excitation.  Since the different resonances correspond to excitation of the different velocity classes this power depletion varies for the different resonance peaks.  This effect was taken into account in the calculation.  For a given resonant velocity class, $v$, the number of atoms resonant with the first stage of the transition was calculated using a Voigt profile based on the Doppler distribution and number density derived from the saturated vapor pressure.  The depletion of the resonant intensity for the first stage of the transition was then modeled as
\begin{align}
  I_1 = I_{10} \, e^{-\sigma_0\, \xi\, n(v) \, L} ,
\end{align}
where, $I_{10}$ is the undepleted intensity, $\sigma_0$ is the resonant scattering cross section for the transition of interest, $\xi$ is the isotopic abundance, $n\prn{v}$ is the density of atoms in the resonant velocity class, and $L$ is an effective scattering length.  The effective scattering length was adjusted to provide good qualitative agreement with the experimental data.  Power depletion of the light resonant with the second stage of the transition is unlikely to be significant since the population of the intermediate state is much less than the population of the ground state.

\section{Experimental Setup}\label{sec expSetup}

Data were collected by use of two different experimental setups.  One of them utilized a frequency comb based on a Ti:Sapphire laser located at Oberlin College.  The other utilized a frequency comb based on a fiber laser located at California State University -- East Bay.

\subsection{The Ti:Sapphire-Comb Experiment}
The frequency comb based on the Ti:Sapphire mode-locked laser is similar to the ``Standard Ti:Sapphire" resonators discussed in Ref.\ \cite{bartels05}.  The mirrors of the cavity consist of two mirrors with a radius of curvature of $30$ cm and two plane mirrors.  The group-velocity dispersion (GVD) of the curved mirrors is compensated with a net negative group-velocity dispersion of $-100$~fs$^2$ over a bandwidth of $620$-$1050$~nm.  One of the plane mirrors also has negative group velocity dispersion of $-40$~fs$^2$.  The other plane mirror is $99\%$ reflective and serves as the output coupler.  The Ti:Sapphire crystal has an absorbance of $\alpha = 5.65$ at 514~nm and an optical path length of 1.5~mm.  The repetition rate of the laser is $\approx 925$ MHz.  The Ti:Sapphire resonator is pumped with 5.5 W of light from a diode-pumped frequency-doubled neodymium vanadate laser at 532 nm.  The spectral bandwidth of the output of the Ti:Sapphire laser is $\approx 30$~nm, centered at $\approx 780$~nm with an output power of $\approx 550$ mW.  This light is coupled into $\approx 30$~cm of a nonlinear photonic crystal fiber with a zero GVD wavelength at $790$~nm.  The light is spectrally broadened by the fiber to span $\approx 500$-$1100$~nm.  Approximately 200~mW of light is coupled through the photonic crystal fiber.  The output of the fiber is split according to its spectral region by use of a short-pass interference filter.  The spectral regions from 500-650~nm and from 1000-1100 nm are used for the stabilization of the carrier envelop offset frequency, $f_0$, and the repetition rate, $f_r$, of the optical frequency comb, while the remaining spectral region can used for spectroscopy.

The carrier envelope offset frequency is detected via the self-referencing technique described in Ref.\ \cite{jones00}.  The infrared light near 1060~nm is doubled using a periodically poled lithium niobate crystal with a length of 1~mm that is thermally stabilized to 381~K.  The doubled infrared light and the green light that is directly produced in the fiber are filtered by use of an interference filter centered at 530~nm with a pass band of 30~nm and detected with an amplified photodiode with a bandwidth of 2~GHz.  The interference of the doubled infrared light with the directly generated green light results in a radio-frequency signal at the carrier-envelope offset frequency.  The signal-to-noise ratio of $f_0$ is typically 40~dB in a 300~kHz resolution bandwidth. The carrier-envelop offset frequency is phase-locked to a signal generator using standard phase-locking techniques (see e.g.\ Ref.\ \cite{stalnaker07}).  The line width of the offset frequency is typically $<500$~kHz.  The repetition rate is detected with the same detector that is used to detect $f_0$.  The output of the photodiode is split and filtered to select the repetition rate.  The repetition rate is phase stabilized to a second signal generator using an rf mixer.  The signal generators are referenced to a rubidium atomic clock that is steered to the global position system (GPS) resulting in a fractional frequency uncertainty of $\approx 10^{-11}$ in 1 second.

The 650-1000 nm light generated from the nonlinear photonic crystal fiber is passed through an interference filter to select the wavelengths of interest.  For excitation through the $\textrm{5P}_{3/2}$ intermediate state the light is passed through an interference filter centered at 780~nm with a 20~nm pass band.  For excitation through the $\textrm{5P}_{1/2}$ state the light is passed through an interference filter that transmitted light from $\approx 785$-860~nm.  This spectral range allows for excitation of the $\textrm{5S}_{1/2}\rightarrow \textrm{5P}_{1/2}\rightarrow \textrm{5D}_{3/2}$ and the $\textrm{5S}_{1/2}\rightarrow \textrm{5P}_{3/2}\rightarrow \textrm{5D}_{3/2, 5/2}$ transitions.  The light is focused with a $10$-cm lens onto a rubidium vapor cell at $\approx 328$~K.  The vapor cell is a cylinder with a 25~mm diameter and a 75~mm length.  The light is propagated through the curved wall of the vapor cell.  A second lens with a focal length of $10$~cm is used to re-collimated the laser beam.  The light is retroreflected off a mirror and sent back through the vapor cell.  A small fraction of the light is picked off for power monitoring by use of a $8\%$ reflective pellicle beam splitter.

Atoms excited to the $5\textrm{D}_{5/2}$ state decay via cascade through the $6\textrm{P}_{1/2, 3/2}$ states, emitting light at 420 nm.  This fluorescence is detected with a photomultiplier tube with an active area of $0.50\:\textrm{cm}^2$.  An interference filter centered at 420~nm with a pass band of 20~nm and a short-pass filter at 650 nm are placed in front of the photomultiplier tube to reduce background and scattered light.  The photocurrent from the photomultiplier tube is sent to a transimpedance amplifier with a gain of $10^7$~V/A and a bandwidth of 20 kHz.  The output of the amplifier is digitized with a analog-to-digital converter and recorded on a computer.  The repetition rate of the optical frequency comb is stepped in 1~Hz intervals over a region of $\approx 1.5$~kHz and the fluorescence is recorded for $\approx 1$ second at each frequency step.

\subsection{The Fiber-Comb Experiment}

The frequency comb based on the erbium-doped fiber laser is a commercial system from Menlo Systems GmbH (FC1500-250-WG Optical Frequency Synthesizer) consisting of a femtosecond laser system (M-Comb) with repetition rate $f_r \approx 250~{\rm MHz}$ whose output is connected via fiber-optic cables to (1) a fast PIN photodiode that detects the fourth harmonic of $f_r$ (for stabilization/control of $f_r$), (2) an erbium-doped fiber amplifier (M-Phase EDFA) whose output is sent to an $f-2f$ interferometer (XPS-WG 1500) used to measure the carrier-envelope offset frequency $f_0$ (for stabilization/control of $f_0$), and (3) a second erbium-doped fiber amplifier whose output is directed through a second-harmonic generating (SHG) crystal.  The output power after the SHG crystal is typically $\approx 230~{\rm mW}$ centered about 780~nm with a FWHM bandwidth of $\approx 15~{\rm nm}$.  The laser light is then focused into a photonic crystal fiber (PCF) which spectrally broadens the output to span roughly from 600-900~nm.  The power after the PCF is typically $\approx 85~{\rm mW}$.

The fourth harmonic of $f_r$ detected by the fast PIN photodiode is mixed with a 980 MHz signal generated by a dielectric resonator oscillator (DRO) to produce a 20~MHz intermediate frequency, which is counted and further mixed with the $\approx 20~{\rm MHz}$ output of a tunable low frequency direct-digital synthesizer (DDS), which enables adjustment of $f_r$ around its nominal value of $250~{\rm MHz}$.  The DC signal resulting from the second stage of mixing is used in a feedback loop to lock $f_r$ by controlling an intra-cavity piezo in the M-Comb laser.  All synthesizers and counters are referenced to a 10~MHz signal from a GPS-disciplined, ultrastable quartz oscillator (TimeTech Reference Generator) with relative stability better than $5 \times 10^{-12}$ in 1 second.

The offset frequency $f_0$, which is tuned to 20~MHz, is counted and fed into a lock-in detector. The latter is referenced to a 20~MHz signal directly generated from the clock signal by frequency doubling. The lock-in detector output signal is used in a feedback loop to lock $f_0$ by controlling the M-Comb laser pump power.

The laser light spanning 600-900~nm generated by the PCF is then split into two beams with a 50/50 nonpolarizing beamsplitting cube.  The beams pass through 10-nm bandpass interference filters centered near the wavelengths of the transitions of interest.  One of the beams (corresponding to the second stage of the two-photon transition of interest) passes through a chopper wheel which modulates the light at $\approx 250~{\rm Hz}$.  Mirrors direct the two beams so that they are counterpropagating and are focused by antireflection-coated, 10-cm focal length bi-convex lenses at the center of a cylindrical Pyrex cell (25~mm in diameter and 75~mm in length) containing Rb vapor heated to $\approx 323~{\rm K}$.  The beams enter perpendicular to the flat circular cell windows and counter-propagate along the axis of the cell.  The focal point of the beams is at the center of the cell.  Fluorescence from the cascade decay of the upper ${\rm 5D_{5/2} }$ and ${\rm 5D_{3/2} }$ states at 420~nm is monitored by an amplified photomultiplier tube (ThorLABs PMM01, active area diameter = 22~mm) fitted with an interference filter centered at 420~nm with a 10~nm pass band to reduce background light.  The PMT is located to the side of the cell directly viewing the center of the cell where the counter-propagating beams are focused.  Approximate alignment of the counter-propagating beams is achieved by directing the two focused beams through 100~$\mu$m pinholes without the cell in place.  The alignment is subsequently improved by adjusting the position of the lenses to maximize the detected fluorescence signal when $f_r$ is tuned to a resonant value for the two-photon transition.  The output of the amplified PMT is sent to the input of a lock-in amplifier (Signal Recovery Model SR7265) referenced to the chopper wheel.  The demodulated output of the lock-in amplifier (time constant = 2~s) is recorded as $f_r$ is stepped in 0.25~Hz increments every 4~s.


\section{Results and Discussion}\label{sec Results}

\subsection{The Ti:Sapphire Comb Data and Calculations}
\begin{figure}[h]
\centering
\includegraphics[width=3.375in]{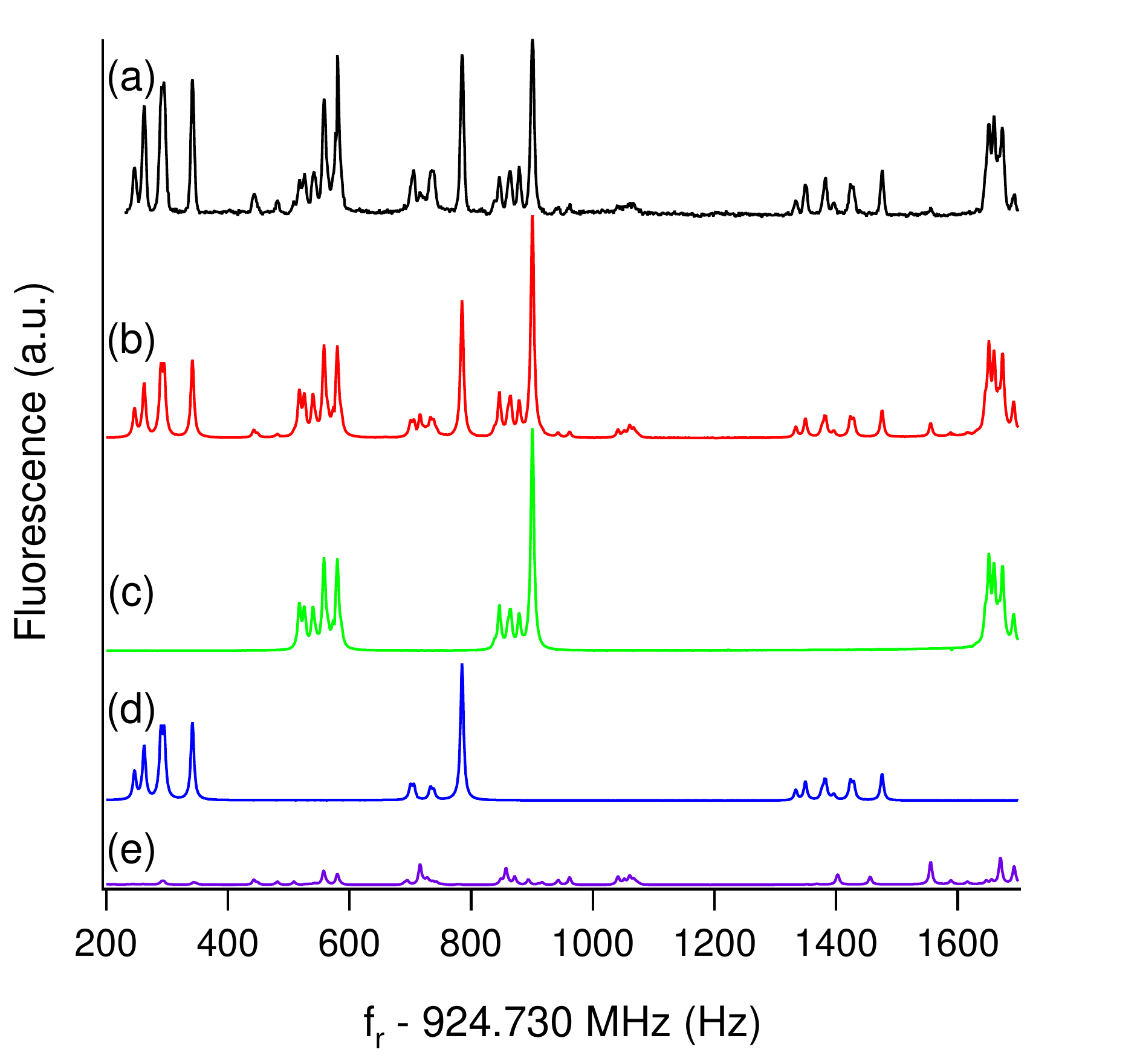}
\caption{Color Online: Trace (a) is the experimental spectrum taken with the Ti:Sapphire comb.  Trace (b) shows the calculated fluorescence spectra for the $\textrm{5S}_{1/2}\rightarrow \textrm{5P}_{1/2} \rightarrow \textrm{5D}_{3/2}$ and $\textrm{5S}_{1/2}\rightarrow \textrm{5P}_{3/2} \rightarrow \textrm{5D}_{3/2,\,5/2}$ transitions.  Traces (c) and (d) show the calculation for the $\textrm{5S}_{1/2}\rightarrow \textrm{5P}_{1/2} \rightarrow \textrm{5D}_{3/2}$ transitions for $^{85}\textrm{Rb}$ and $^{87}\textrm{Rb}$, respectively. Trace (e) shows the calculation for the $\textrm{5S}_{1/2}\rightarrow \textrm{5P}_{3/2} \rightarrow \textrm{5D}_{3/2,5/2}$ transition for both isotopes.  The amplitude of the transitions through the $\textrm{5P}_{3/2}$ state relative to the those through the $\textrm{5P}_{1/2}$ state were adjusted to reflect the differing light intensity transmitted through the filter.}\label{fig OberlinDataD32}
\end{figure}

\begin{figure}[h]
\centering
\includegraphics[width=3.375in]{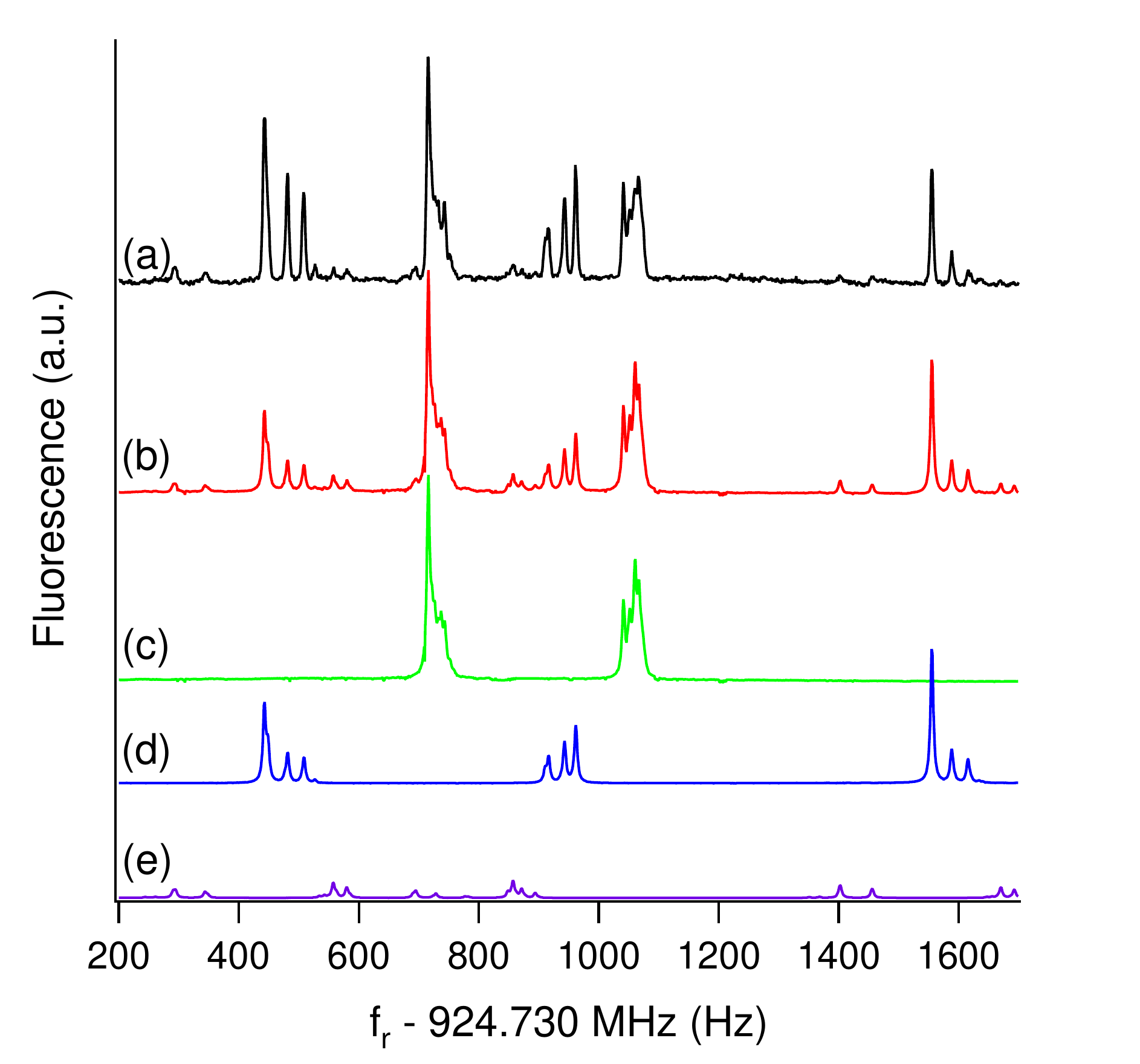}
\caption{Color Online: Trace (a) is the experimental spectrum taken with the Ti:Sapphire comb.  Trace (b) shows the calculated fluorescence spectra for the $\textrm{5S}_{1/2}\rightarrow \textrm{5P}_{3/2} \rightarrow \textrm{5D}_{5/2}$.  Traces (c) and (d) show the calculation for the $\textrm{5S}_{1/2}\rightarrow \textrm{5P}_{3/2} \rightarrow \textrm{5D}_{5/2}$ transitions for $^{85}\textrm{Rb}$ and $^{87}\textrm{Rb}$, respectively. Trace (e) shows the calculation for the $\textrm{5S}_{1/2}\rightarrow \textrm{5P}_{3/2} \rightarrow \textrm{5D}_{3/2}$ transition for both isotopes.  This transition is nine times weaker due to the relative amplitudes of the reduced matrix elements and is also weighted by the branching ratio to the $\textrm{6P}$ states (38\% for the $\textrm{5D}_{3/2}$ state and 35\% for the $\textrm{5D}_{5/2}$ state based on the calculations of Ref.\ \cite{heavens61}).}\label{fig OberlinDataD52}
\end{figure}

The experimental and calculated spectra for the $\textrm{5S}_{1/2}\rightarrow \textrm{5P}_{1/2} \rightarrow \textrm{5D}_{3/2}$ and $\textrm{5S}_{1/2}\rightarrow \textrm{5P}_{3/2} \rightarrow \textrm{5D}_{5/2}$ transitions from the Ti:Sapphire-comb experiment are shown in Figs.\ \ref{fig OberlinDataD32} and \ref{fig OberlinDataD52}.  The data in Fig.\ \ref{fig OberlinDataD32} were taken with an interference filter that had a stronger transmission at 794 nm (resonant wavelength for the $\textrm{5S}_{1/2}\rightarrow \textrm{5P}_{1/2}$ transition) than at 780 nm (resonant wavelength for the $\textrm{5S}\rightarrow \textrm{5P}_{3/2}$ transition).  This results in a spectrum largely consisting of the $\textrm{5S}_{1/2}\rightarrow \textrm{5P}_{1/2} \rightarrow \textrm{5D}_{3/2}$ transitions, with much smaller contributions from the $\textrm{5S}_{1/2}\rightarrow \textrm{5P}_{3/2} \rightarrow \textrm{5D}_{3/2, 5/2}$. transitions.  The data in Fig.\ \ref{fig OberlinDataD52} were taken with an interference filter that did not transmit light at 794 nm so that the spectrum consists entirely of the $\textrm{5S}_{1/2}\rightarrow \textrm{5P}_{3/2} \rightarrow \textrm{5D}_{3/2, 5/2}$ transitions. There is excellent agreement in the peak positions for both sets of data.  The $\textrm{5S}_{1/2}\rightarrow \textrm{5P}_{3/2} \rightarrow \textrm{5D}_{3/2}$ transitions are much smaller than the $\textrm{5S}_{1/2}\rightarrow \textrm{5P}_{3/2} \rightarrow \textrm{5D}_{5/2}$ transitions due to angular momentum considerations.

The amplitude agreement is reasonable, given the assumptions of the model.  We attribute the residual disagreement between the calculated and experimental amplitudes to issues related to the power depletion of the light exciting the first stage of the transition.  The calculation includes this effect but assumes that the counterpropagating beams have common focal points which localizes the position at which the atoms are excited. If the focal positions of the two laser beams do not overlap perfectly, the location where the atoms are excited can vary depending on the relative powers of the two beams.  This can combine with position-dependent fluorescence detection efficiency and affect the relative amplitudes of resonances from different velocity classes.  We note that an effective optical depth of 1.0~cm was used in the calculation, which agrees well with geometrical size of the cell.

The calculations were done with a transit line width of $\frac{\gamma_T}{2\pi} = 4$~MHz. This value is somewhat larger than the estimated line width if $\frac{\gamma_T}{2\pi} = 500$~kHz, based on a laser beam diameter of $\approx 80\:\mu \textrm{m}$.  We attribute this additional broadening to imperfect overlap in the focal points of the counter-propagating beams.  If the foci of the two beams do not overlap completely then there will be atoms for which the wave vectors of the two beams are not antiparallel.  These atoms will have an additional broadening due to the imperfect cancelation of the Doppler shift.  We explored the effect of wave vectors that were slightly misaligned by including a misalignment term in the calculation.  The calculation reproduced broadening at the same level we observe in the experiment for values that are consistent with our limits of how well the foci of the two beams overlap.  The calculations were done assuming linearly polarized light.  For the data, the polarization of the the light was not controlled.  However, data were taken with linearly polarized light and no differences in the spectra were observed.

Trace (c) of Fig.\ \ref{fig OberlinDataD32} shows the calculated spectrum for $\textrm{5S}_{1/2}\rightarrow \textrm{5P}_{1/2}\rightarrow \textrm{5D}_{3/2}$ transition for $^{85}\textrm{Rb}$.  The spectrum consists of three distinct groups.  The first group on the left is due to excitation from the $F=3\rightarrow F^{\prime \prime}$ transitions, while the second grouping corresponds to the $F=2\rightarrow F^{\prime \prime}$ transitions.  The last group at the high end of the spectrum corresponds again to the excitation of the $F=3\rightarrow F^{\prime \prime}$ transitions, but with a new total mode number equal to one fewer than the mode number corresponding to the first group.  The frequency spacing between the first and third group corresponds the the repetition interval predicted by Eq.\ \eqref{eq repInterval}, $\Delta f_r \approx 1.1$~kHz.  The relative amplitudes of the groups, as well as the amplitudes of the peaks within each group, vary significantly due to the different velocity classes that are excited by the different modes of the frequency comb.  Such features are also evident in the $\textrm{5S}_{1/2}\rightarrow \textrm{5P}_{3/2}\rightarrow \textrm{5D}_{5/2}$ transitions (Fig.\ \ref{fig OberlinDataD52}).

In both spectra the hyperfine states of the ground and excited states are clearly resolved while the intermediate hyperfine structure is not.  This is in contrast to the data in Ref.\ \cite{stalnaker10} where the transitions through different intermediate states gave distinct peaks.  This difference is a result of the relatively close energies of the two resonant photons.

\subsection{The Fiber Comb Data and Calculations}

\begin{figure}[h]
\centering
\includegraphics[width=3.375in]{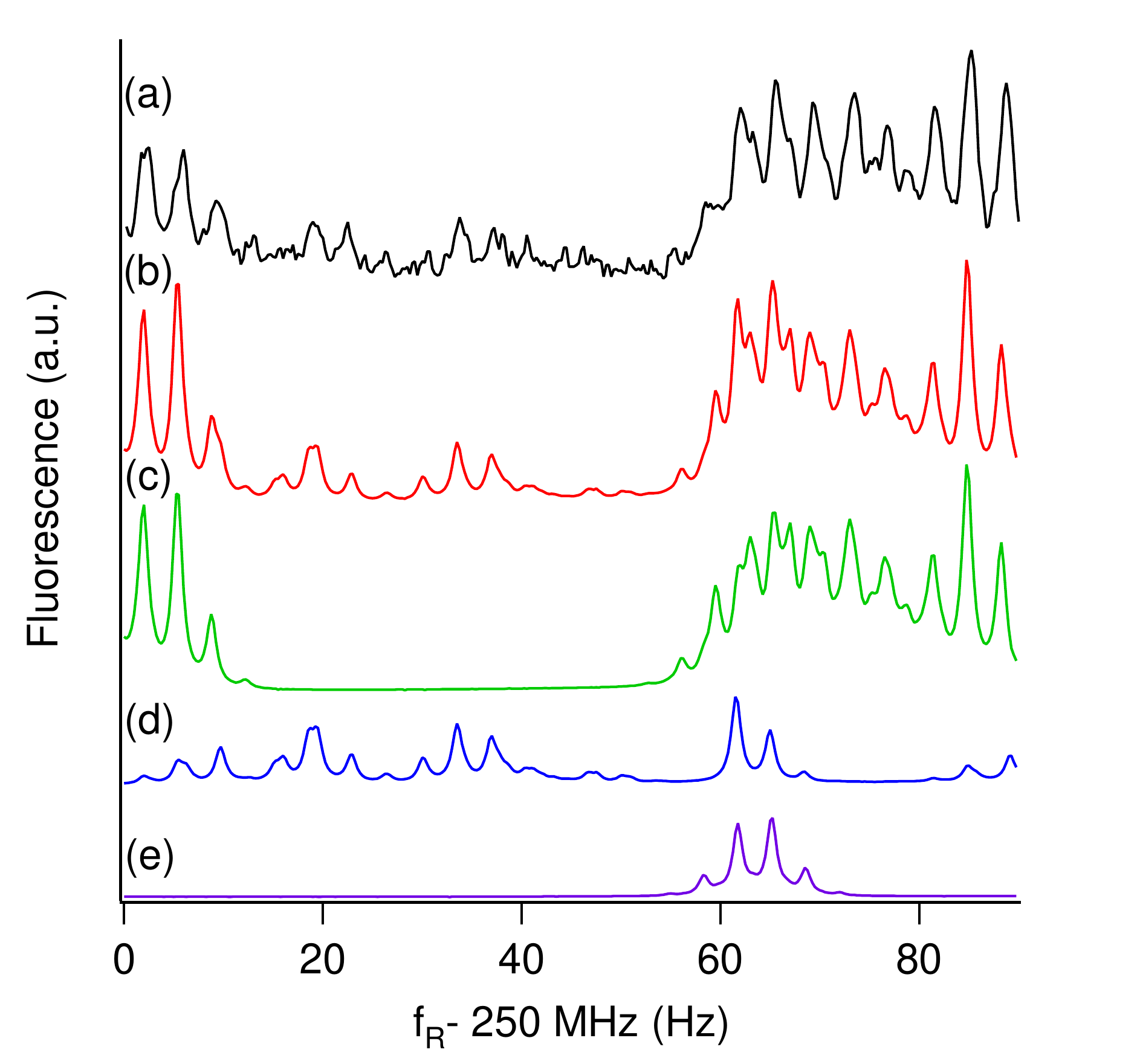}
\caption{Color Online: Trace (a) is the experimental spectrum taken with the fiber comb.  Trace (b) shows the calculated fluorescence spectra for the $\textrm{5S}_{1/2}\rightarrow \textrm{5P}_{1/2} \rightarrow \textrm{5D}_{3/2}$ transitions.  Traces (c) and (d) show the calculation for the $\textrm{5S}_{1/2}\rightarrow \textrm{5P}_{1/2} \rightarrow \textrm{5D}_{3/2}$ transitions for $^{85}\textrm{Rb}$ and $^{87}\textrm{Rb}$, respectively. Trace (e) shows the calculation for the $F=3\rightarrow F^{\prime\prime}=2$ transition for $^{85}\textrm{Rb}$.  Several distinct resonances occur for different mode combinations, each of which have the same total mode number. }\label{fig CSEBDataD32}
\end{figure}

\begin{figure}[h]
\centering
\includegraphics[width=3.375in]{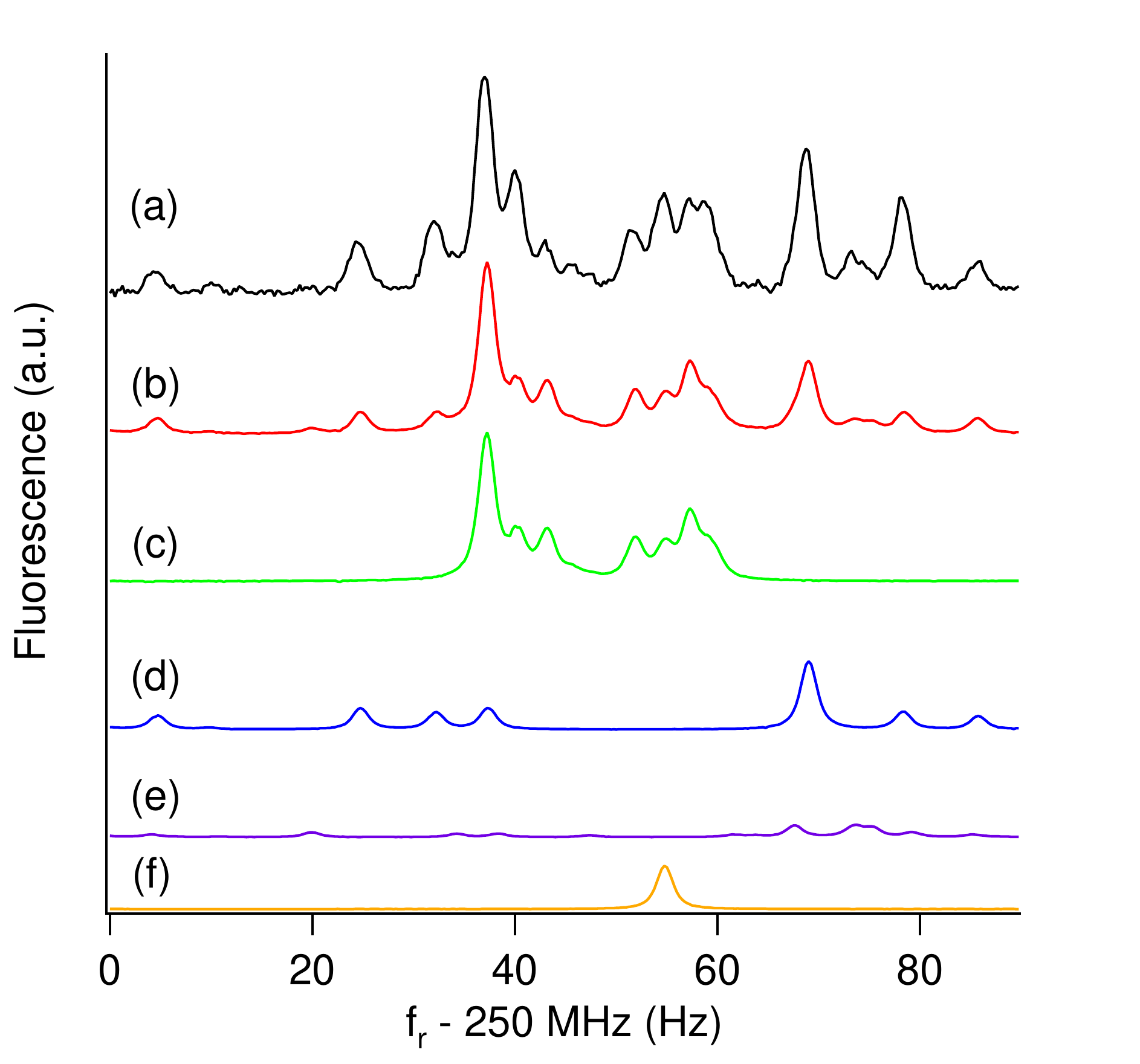}
\caption{Color Online: Trace (a) is the experimental spectrum taken with the fiber comb.  Trace (b) shows the calculated fluorescence spectra for the $\textrm{5S}_{1/2}\rightarrow \textrm{5P}_{3/2} \rightarrow \textrm{5D}_{3/2,5/2}$ transitions.  Traces (c) and (d) show the calculation for the $\textrm{5S}_{1/2}\rightarrow \textrm{5P}_{3/2} \rightarrow \textrm{5D}_{5/2}$ transitions for $^{85}\textrm{Rb}$ and $^{87}\textrm{Rb}$, respectively.  Trace (e) shows the calculation for the $\textrm{5S}_{1/2}\rightarrow \textrm{5P}_{3/2} \rightarrow \textrm{5D}_{3/2}$ transition for both isotopes.  This transition is nine times weaker due to the relative amplitudes of the reduced matrix elements and is also weighted by the branching ratio to the $\textrm{6P}$ states (38\% for the $\textrm{5D}_{3/2}$ state and 35\% for the $\textrm{5D}_{5/2}$ state based on the calculations of Ref.\ \cite{heavens61}).  Trace (f) shows the calculation for the $F=2\rightarrow F^{\prime\prime}=3$ transition for $^{85}\textrm{Rb}$.}\label{fig CSEBDataD52}
\end{figure}

The experimental and calculated spectra for the $\textrm{5S}_{1/2}\rightarrow \textrm{5P}_{1/2} \rightarrow \textrm{5D}_{3/2}$ and $\textrm{5S}_{1/2}\rightarrow \textrm{5P}_{3/2} \rightarrow \textrm{5D}_{3/2,5/2}$ transitions from the fiber-comb experiment are shown in Figs.\ \ref{fig CSEBDataD32} and \ref{fig CSEBDataD52}, respectively.  For the spectra shown in Figs.\ \ref{fig CSEBDataD32} and \ref{fig CSEBDataD52}, interference filters with 10-nm bandwidths were employed.  These interference filters provided sufficient wavelength selectivity so that in Fig.\ \ref{fig CSEBDataD32} only excitation through the $\textrm{5S}_{1/2} \rightarrow \textrm{5P}_{1/2}$ pathway is observed and in Fig.\ \ref{fig CSEBDataD52} only excitation through the $\textrm{5S}_{1/2} \rightarrow \textrm{5P}_{3/2}$ is observed.

As with the Ti:Sapphire-comb experiment, the relative amplitudes of the different resonance peaks differ slightly between the calculated and experimental spectra.  The position of the peaks agree well for the $\textrm{5S}_{1/2}\rightarrow \textrm{5P}_{3/2} \rightarrow \textrm{5D}_{5/2}$ spectrum.  However, for the $\textrm{5S}_{1/2}\rightarrow \textrm{5P}_{1/2} \rightarrow \textrm{5D}_{3/2}$ spectrum there is some discrepancy at the 0.5~Hz level. We attribute this to the increased overlap and complexity of the spectrum.   Shifts in the apparent positions of the transition resonances will occur as a result of the discrepancies in relative amplitudes between the calculated and experimental spectra.

Power depletion of the light corresponding to excitation of the first stage of the transition was not found to improve the agreement between the data and calculation and was not included.  This may be a result of a lower atomic density for the data taken with the fiber comb compared to that for the Ti:Sapphire comb experiment.  The calculations for the $\textrm{5S}_{1/2}\rightarrow \textrm{5P}_{1/2} \rightarrow \textrm{5D}_{3/2}$ transitions were done with a transit line width of $\frac{\gamma_T}{2\pi} = 2.7$~MHz.  Those for the $\textrm{5S}_{1/2}\rightarrow \textrm{5P}_{1/2} \rightarrow \textrm{5D}_{3/2}$ transitions were done with a transit line width of $\frac{\gamma_T}{2\pi} = 4$~MHz.  These values are in reasonable agreement with the estimated based on a laser beam diameter of $15\:\mu \textrm{m}$, although additional broadening due to imperfect overlap of the foci of the counter propagating laser beams may have contributed additional broadening.   The calculations were done assuming linearly polarized light.  For the data shown, the polarization of the the light was not controlled.  However, data were taken with linearly polarized light for the transitions excited through the $\textrm{5P}_{3/2}$ state and these data did not display any differences from the data where the polarization was not controlled.

For a comb with a repetition rate of 250~MHz, $f_r$ must be scanned by $\Delta f_r \approx 80$~Hz in order for a given transition to recur in the spectrum.  As can be seen from Eq.\ \eqref{eq repInterval}, this is a factor of the ratio of the repetition rates squared smaller than the repetition interval for the Ti:Sapphire comb.  The widths of the resonance peaks decrease by a single power of the ratio of the repetition rates.  As a result, there is more overlap of the peaks arising from different transitions.  This is clearly observable in the comparison of $\textrm{5S}_{1/2}\rightarrow \textrm{5P}_{3/2}\rightarrow \textrm{5D}_{5/2}$ transitions for the different combs, Figs.\ \ref{fig OberlinDataD52} and \ref{fig CSEBDataD52}.  However, in the case of the $\textrm{5S}_{1/2}\rightarrow \textrm{5P}_{3/2}\rightarrow \textrm{5D}_{5/2}$ transition the individual peaks are still resolvable in most cases.  Trace (f) shows the calculated spectrum for a single hyperfine transition, in this case the $F=2\rightarrow F^{\prime\prime}=3$ for $^{85}\textrm{Rb}$.  As expected, there is only one resonant peak in the spectrum for this transition, a result of the near degeneracy of the two photons involved in the transition.

As described above, the situation is quite different for the $\textrm{5S}_{1/2}\rightarrow \textrm{5P}_{1/2}\rightarrow \textrm{5D}_{3/2}$ transitions.  Trace (e) of Fig.~\ref{fig CSEBDataD32} shows the calculated spectrum for a single hyperfine transition of $^{85}\textrm{Rb}$ ($F=3\rightarrow F^{\prime\prime}=2$).  There are multiple distinct velocity classes that contribute to the fluorescence signal and the resonance for each velocity class occurs at a different repetition rate.  The result is multiple distinct peaks in the fluorescence signal.  This results in a significantly more complicated fluorescence spectrum where there are numerous peaks overlapping peaks.

\section{Conclusions}

We have presented a comprehensive investigation of velocity-selective two-photon direct frequency comb spectroscopy in atomic Rb.  We have experimentally and theoretically demonstrated the effect of the repetition rate of a frequency comb on the two-photon excitation rate.  The energy level structure of Rb also allowed us to explore the effect of the energy of the intermediate state on the two-photon excitation.  The energy of the intermediate state is particularly important in the case where the repetition rate of the frequency comb is less than the Doppler width of the resonance corresponding to the first stage of the transition.

This investigation demonstrates the benefits and challenges of direct-frequency-comb spectroscopy.  While there is significant advantage in the wavelength versatility of the frequency comb, the details of resulting spectra can display complicated features resulting from the presence of the numerous frequencies.  This effect is particularly pronounced in the case where the repetition rate of the comb is less than the Doppler distribution of the atoms.  The presence of multiple resonances corresponding to excitation of a two-photon resonance from comb light with different mode numbers results in a dense spectrum.  This effect will likely complicate any effort to make quantitative measurements of transition frequencies for atoms with a multiple hyperfine transitions.

\section{Acknowledgements}

The authors would like to acknowledge Scott Diddams for assistance with the Ti:Sapphire oscillator, Lee Sherry and William Striegl for early contributions to the Oberlin frequency comb experiment, and Bill Marton for help with the construction of the apparatus.  The Oberlin frequency comb experimented benefited from funding from the National Institute of Standards and Technology Precision Measurements Grant.  The California State University - East Bay frequency comb experiment was supported by the National Science Foundation under Awards PHY-0958749 and PHY-0969666.

\end{document}